# Can EGRET's Gamma Ray Sources > 100 MeV Be Seen with Single Secondary Cosmic Muons from Gammas > 30 GeV?

J. Carpenter, S. Desch, T. F. Lin, J. Poirier, and A. Roesch

*Physics Department, 225 NSH, University of Notre Dame, Notre Dame, IN 46556, USA*

## Abstract

The high energy gramma ray sources provided by experiment EGRET on the Compton Gamma Ray Observatory are examined, testing Project GRAND's ability to detect primary gamma rays by means of studying secondary muons. There is ~1.5% chance that a 100 GeV primary cosmic gamma ray will produce a muon at detection level (Fasso & Poirier, 1999), a probability which increases with increasing gamma energy. Project GRAND has 80 square meters of muon detector which identifies secondary muons > 0.1 GeV and measures their angles to 0.26 deg (projected angle in the XZ and YZ planes). Data taken during the last two years are analyzed. A table of EGRET's gamma ray sources is examined (The Third EGRET Catalog of High Energy Gamma Ray Sources (EGRET webpage, 1999). EGRET's flux (> 0.1 GeV), angular information, and spectral index were extrapolated to GRAND's energy region (>30 GeV). Then the geometrical acceptance was calculated for each of these events. Thus, GRAND's sensitivity to each of EGRET's sources was predicted. A product of extrapolated flux and GRAND's detection sensitivity yields an overall parameter indicating GRAND's relative sensitivity to each source, assuming an energy dependence given by EGRET's spectral index. The 14 sources with the best extrapolated detection efficiency were selected to be examined with the data.

## 1. Introduction:

Secondary cosmic ray muons at the surface of the earth result mostly from the interaction of hadronic primary cosmic rays with the atmosphere. Primary cosmic rays are mostly hadronic, but a small fraction are gamma rays. However, the gamma rays are more interesting because they have zero charge and so are not deflected by electromagnetic fields in space. Therefore, gamma rays can be traced back to their origin in space. Cosmic gamma rays interact with the atmosphere and produce electron and positron pairs. About 1.5% of gamma ray interactions at 100 GeV produce pions which then decay into muons (Fasso & Poirier, 1999). By looking at these secondary muons, Project GRAND thus has a sensitivity to these primary gamma rays. EGRET lists available point sources detected in the 100 MeV range on the website:

http://cossc.gsfc.nasa.gov/cossc/egret/egretform.html.

By analyzing these secondary muon tracks in Project GRAND's single track data collected over the past two years, it can be determined if those sources were also detected at this higher energy range in this data. Project GRAND detects cosmic rays in an energy range of 10-300 GeV utilizing secondary single muon tracks whereas EGRET detects cosmic gamma rays in an energy range of 20 MeV to 30 GeV. By gathering data at a significantly higher energy, more properties of these primary cosmic ray sources could be learned, especially their spectral index or how rapidly they turn off with increasing energy.

## 2. Experiment:

Project GRAND (Gamma Ray Astrophysics at Notre Dame) is a 100 x 100 meter detector array of 64 huts located just north of the University of Notre Dame approximately 150 km east of Chicago at 86°W by 42°N. Each square hut is 2.4 m wide and 1.5 m in height. Each hut contains four proportional wire chambers (PWCs) each containing two planes. A 50 mm steel plate is positioned between the third and fourth PWC. About 96% of the muons pass through the steel plate without producing high energy delta-ray electrons which would create additional hits in the fourth PWC (Kochocki et al., 1991). This allows the huts to detect muons with a 96% efficiency and determine the angles of their tracks with an average of 0.26 degrees (projected). The secondary muons that are detected are the result of interactions of the primary cosmic rays with the atmosphere producing pions, which then decay to muons. The pions are produced at a small angle relative to the primary gamma ray, the decay muons are at a small angle relative to the pion, and the muons are then bent by the earth's magnetic field and scattered in the earth's atmosphere resulting in an effective resolution for the primary cosmic ray of ± 5° (which includes all these effects).

## 3. Data Analysis:

For each sidereal day, data were taken from the past two years for 360 degrees of right ascension and 110 degrees of declination from -20 to 90 degrees. Only data that cover a complete sidereal day were used. These data were then binned into one degree by one degree squares. The data were then projected onto the right ascension axis creating a histogram. The maximum and minimum values for this projection were found. If these values differed by more than ± 5% from the average of that data, then all the data from this sidereal day were discarded. A sidereal day's data which were not uniform could be a result of artificial causes, for example, some percentage of huts malfunctioning or being turned off for a period of time for repairs. The remaining smooth data were added to get a complete picture of the sky visible to GRAND.

EGRET's website provided a list of point sources which they detected. Only those with a positive declination were chosen because Project GRAND can not see negative declinations efficiently since these angles are near the local horizon. It was then useful to find the sources that had the greatest probability of being detected by GRAND. First, Project GRAND's acceptance rate times the angle dependence of the muon flux as a function of zenith angle was parameterized by this equation:

$$\text{Acceptance} = (X / L) * \cos(\theta) * \cos^2(\theta)$$

**Figure 1:** Geometry to calculate Project GRAND's geometric acceptance as a function of angle from zenith, $\theta$.

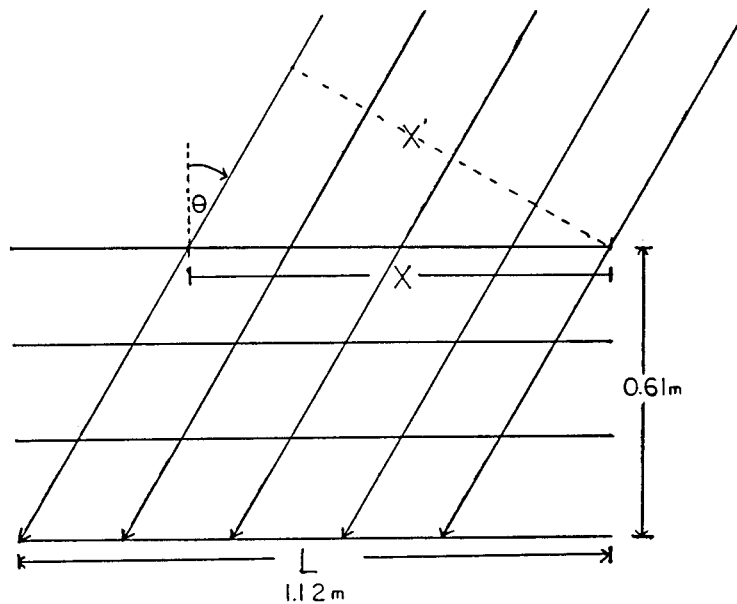

The ratio of X to L from Fig. 1 represents the reduced length of the top PWC plane for a muon at angle $\theta$ relative to zenith. This ratio is multiplied by $\cos(\theta)$ to project the length X normal to the muon track, X'.

The angular distribution of muons from zenith angle is a function of $\cos^2(\theta)$ (Particle Data Group, 1998). Project GRAND's highest acceptance occurs when the muon track is near the zenith angle which, for GRAND's location on the earth, is 42° declination. From EGRET's table, the flux and the spectral index (gamma) were used to extrapolate EGRET's results to GRAND's energy range to determine the expected flux rate as seen by GRAND:

$$\text{GRAND's flux} = \text{EGRET's flux} * 300^{-(\gamma - 1.0)}$$

The factor of 300 results from the energy ratio of EGRET's energy > 0.1 GeV to Project GRAND's energy > 30 GeV. EGRET's flux projected to GRAND's energy was then multiplied with the detector acceptance for that declination angle. From these extrapolated results, the top 14 sources with the highest predicted extrapolated flux and acceptance were then chosen. Data in a region ± 5° in declination and ± 5° / cos(δ) in RA around the predicted source were analyzed. The width of the right ascension angle decreases with increasing declination and so is divided by cos(δ) to have a greater span of right ascension angles and retain the ± 5° resolution. The neighboring regions of ± 5° / cos(δ) in RA and ± 5° in declination were added together and divided by two to find the background for the source region. Neighboring regions in declination were not considered for this background calculation because the data vary rapidly with declination (Poirier et al., 1997). This background value was subtracted from the source region to find the signal. The error on this signal was found by adding, in quadrature, the statistical errors on the source point region and the background region. Dividing the signal by the error on the signal gives the amount a possible signal is above background (sig/dsig). The results are listed in Table 1. This table contains the declination in degrees (DEC), the right ascension in degrees (RA), GRAND's angular acceptance (acpt) (which includes the $\cos^2\theta$ muon flux variance with zenith angle), the counts of muons (sig), its error (dsig), the background count in millions (back), the signal divided by background in parts per thousand (flux), and the signal over the error in the signal (sig/dsig).

Table 1: Project GRAND's detection of EGRET sources.

| EGRET no. | DEC | RA | acpt | sig | dsig | back | flux | sig/dsig |
|---|---|---|---|---|---|---|---|---|
| EG J0237+1635 | 17 | 40 | 0.55 | -21801 | 15459 | 159.4 | -0.1 | -1.4 |
| EG J0433+2908 | 30 | 69 | 0.83 | -24579 | 19325 | 249.0 | -0.1 | -1.3 |
| EG J0534+2200 | 23 | 84 | 0.69 | -24684 | 17536 | 205.0 | -0.1 | -1.4 |
| EG J0617+2238 | 23 | 95 | 0.69 | 6995 | 17532 | 204.9 | 0.0 | 0.4 |
| EG J0633+1751 | 18 | 99 | 0.58 | 12459 | 15812 | 166.7 | 0.1 | 0.8 |
| EG J1104+3809 | 39 | 167 | 0.97 | -55882 | 21909 | 320.1 | -0.2 | -2.6 |
| EG J1200+2847 | 29 | 181 | 0.81 | -46166 | 19266 | 247.5 | -0.2 | -2.4 |
| EG J1222+2841 | 29 | 186 | 0.81 | -35673 | 19268 | 247.5 | -0.1 | -1.9 |
| EG J1635+3813 | 39 | 249 | 0.97 | 29948 | 21963 | 321.6 | 0.1 | 1.4 |
| EG J1835+5918 | 60 | 279 | 0.71 | 61942 | 20783 | 287.9 | 0.2 | 3.0 |
| EG J1958+2909 | 30 | 300 | 0.83 | 17874 | 19366 | 250.0 | 0.1 | 0.9 |
| EG J2020+4017 | 41 | 306 | 0.99 | 45968 | 21944 | 321.0 | 0.1 | 2.1 |
| EG J2021+3716 | 38 | 306 | 0.95 | 35815 | 21988 | 322.3 | 0.1 | 1.6 |
| EG J2033+4118 | 42 | 309 | 1.00 | 36149 | 21839 | 318.0 | 0.1 | 1.7 |

## 4. Conclusions:

As can be seen from the data, five of the points are greater than one sigma from background, two are above two sigma from background, and one of the points is at or above three sigma from background. From statistics the expected counts were 2.2 events for $\geq +1\sigma$, 0.32 events for $\geq +2\sigma$, and 0.02 events for $\geq +3\sigma$. Statistically, the probability of detecting $\geq 5$ events at $\geq +1\sigma$ is 5.8%, detecting $\geq 2$ events for $\geq +2\sigma$ is 4%, and detecting $\geq 1$ event at $\geq +3\sigma$ is 2.2%. Eight of the points had an abundance and six of the points had a deficiency. The one event at $+3\sigma$ does not appear sufficiently strong to support a positive detection.

This research is presently being funded through grants from the University of Notre Dame and private donations. The National Science Foundation participated in GRAND's construction.

## References

EGRET Catalog: http://cossc.gsfc.nasa.gov/cossc/egret/egretform.html.
Fasso, A. & Poirier J. 1999, Proc of 26th ICRC (Salt Lake City) HE3.2.20
Kochocki, J. et al. 1991, Proc of 22nd ICRC (Dublin) 4, 425
Particle Data Group, Caso, C. et al. 1998, The European Physical Journal C3 1
Poirier, J. et al. 1997, Proc of 25th ICRC (Durban) 6, 369